\documentclass[journal]{IEEEtran}
\usepackage{comment}
\usepackage{graphicx}
\usepackage{amsmath,amsfonts,amssymb}
\usepackage{array}
\usepackage{upgreek}
\begin{document}
\title{Realizing a Robust, Reconfigurable Active Quenching Design for Multiple Types of Single-Photon Avalanche Detectors}
%
%
%

\author{Subash Sachidananda, Prithvi Gundlapalli, Victor Leong, Soe Moe Thar, Leonid Krivitsky and Alexander Ling
\thanks{S. Sachidananda and S. Moe Thar are with the Centre for Quantum Technologies, National University of Singapore, 3 Science Drive 2, Singapore 117543 (e-mail: subash.jies@gmail.com).}
\thanks{A. Ling is with the Centre for Quantum Technologies and Department of Physics, National University of Singapore, 3 Science Drive 2, Singapore 117543 (e-mail: alexander.ling@nus.edu.sg).}
\thanks{P. Gundlapalli, V. Leong and L. Krivitsky are with the Institute of Materials Research and Engineering, Agency for Science, Technology and Research (A*STAR), Singapore 138634 (email: victor\_leong@imre.a-star.edu.sg).}
\thanks{Manuscript received MMM DD, YYYY; revised MMM DD, YYYY.}}

%
%

%

\maketitle

\begin{abstract}
Most active quench circuits used for single-photon avalanche photodetectors (APDs) 
are designed either with discrete components which lack the flexibility of dynamically changing the control parameters, 
or with custom ASICs which require a long development time and high cost. 
As an alternative, we present a reconfigurable and robust hybrid design implemented using a System-on-Chip (SoC), 
which integrates both an FPGA and a microcontroller. 
We take advantage of the FPGA’s speed and reconfiguration capabilities to vary the quench and reset parameters dynamically over a large range, 
thus allowing our system to operate a variety of APDs without changing the design. 
The microcontroller enables the remote adjustment of control parameters and calibration of APDs in the field. 
The ruggedized design uses components with space heritage, 
thus making it suitable for space-based applications in the fields of telecommunications and quantum key distribution (QKD). 
We demonstrate our circuit by operating a commercial APD cooled to -20°C with 
a deadtime of 35ns while maintaining the after-pulsing probability at close to 3$\%$. 
We also showcase its versatility by operating custom-fabricated chip-scale APDs, 
which paves the way for automated wafer-scale characterization.
\end{abstract}

\begin{IEEEkeywords}
single photon avalanche detectors , System on chip SoC, FPGA, active quenching, chip-scale APD, reconfigurable quenching circuit, automated wafer-scale testing.
\end{IEEEkeywords}

\section{INTRODUCTION}
\IEEEPARstart{S}{ingle}-photon detection is used widely in quantum technologies. 
The most popular method for detecting single photons is to use Avalanche Photodetectors (APDs) due to their speed, high efficiency, and the ability to operate at or near room temperature.
When operated in the Geiger mode, the absorption of a single photon triggers a macroscopic current due to an avalanche multiplication process. 
This avalanche current needs to be stopped quickly, and the APD restored to its nominal state, before it can detect the next photon. 
The simplest way to achieve this is with a $passively\ quenched$ circuit~\cite{Cova96, wincomparator} wherein the avalanche multiplication process of the diode is stopped by using a large ballast resistor in series with the diode. 
However this method is slow and thus limits the detector performance. 
An improvement is the $passive\ quench\ active\ reset$ (PQAR) circuit~\cite{pqar}, which also uses a series ballast resistor to stop the avalanche, but actively restores the APD to its nominal state within a short, well-defined duration. 
However, its drawback is that the APD after-pulsing probability goes above acceptable bounds if the reset times are too short. 
$Active\ quench$ circuits~\cite{Gallivanoni10} provide fast response and reset times while limiting the after-pulsing probability,
and thus are the most suitable choice for high photon-rate applications.
They operate in two phases upon the onset of an avalanche: 
1) $quench$ --- reduce the APD bias voltage below the breakdown voltage, and 
2) $reset$ --- restore the bias to above the breakdown voltage while bypassing the ballast resistor. 
After the reset phase, the ballast resistor is reconnected and the APD returns to the nominal state. The duration of the whole process is typically a few nanoseconds.


In many existing active quenching circuit designs, the generation of the quench and reset signals is done using only discrete hardware components~\cite{Ghioni96, Stipcevic09, Stipcevic17}. 
Such designs are cheaper to manufacture, but critical parameters such as the quench and reset times are fixed and cannot be changed dynamically. 
Tuning these parameters dynamically can compensate for deviations in the APD performance due to factors such as temperature variation. 
On the other hand, monolithic integrated active quench circuits~\cite{zappa2000, Acconcia16, 37ps-aq, Ceccarelli2019, Zimmerman5ns} offer the flexibility to vary parameters dynamically while also allowing for very high count rates and detection efficiencies. 
However, fabricating custom ICs is a complex process requiring long development cycles and a very high manufacturing cost. 
\begin{figure*} [!t]
\centering
\includegraphics[width=\textwidth]{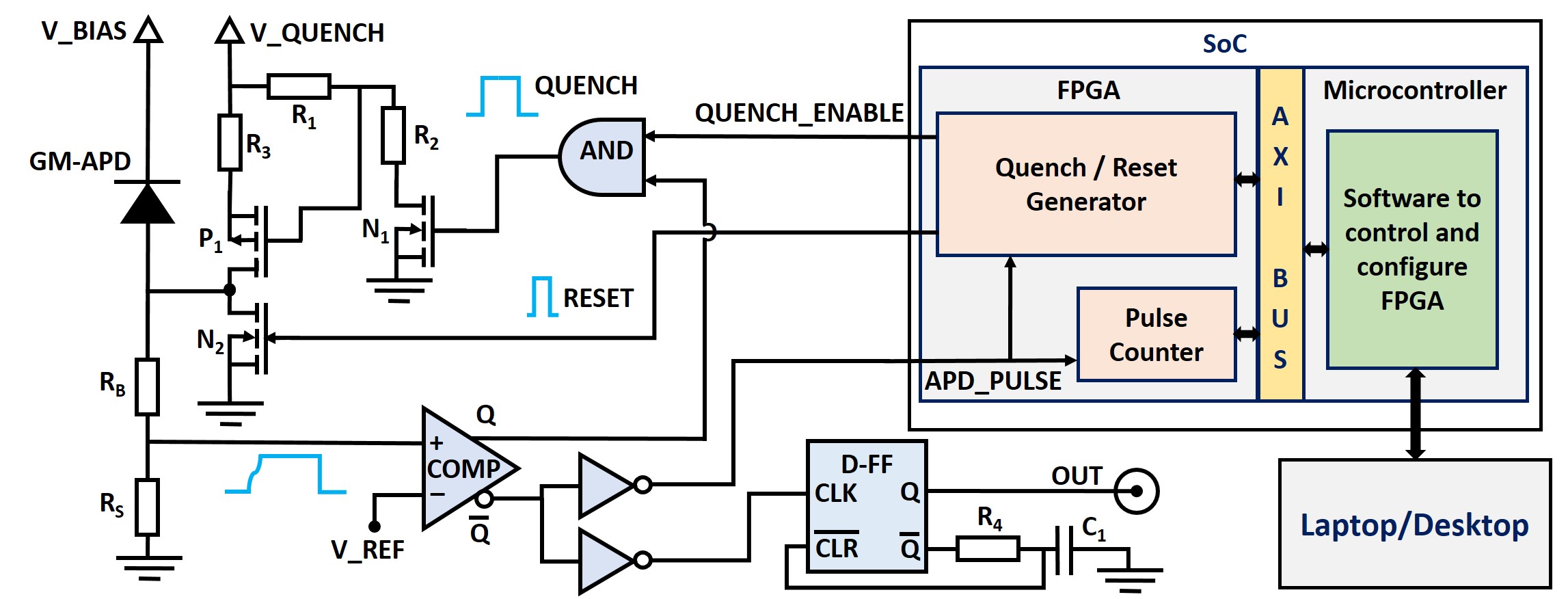}
\caption[] 
{ \label{aq-design2} 
Simplified circuit diagram of the active quench system with a Geiger-mode APD (GM-APD).
The symbols are described in detail in the main text.}
\end{figure*}

In this paper we demonstrate a hybrid design for the active quenching of Geiger-mode APDs 
based on a commercially available System-on-Chip (SoC)~\cite{zedboard} which has both an FPGA and a micro-controller integrated on the same chip. 
The control logic is implemented on the FPGA, allowing for dynamic control of active quench parameters such as quench and reset times. 
The microcontroller interfaces with a PC and provides remote access to the FPGA, thus enabling operations such as re-calibration of deployed APDs in the field. 
Only a few discrete components are placed close to the APD itself, which allows the SoC to be placed farther away without compromising the active quench performance. 
Our results show that our system can achieve a deadtime of 35\,ns with the APD located $\approx$15\,cm away from the SoC, which would allow for count rates of $>$28\,Mcps. 
Although our system is mainly designed for active quenching, we show how it can also be reconfigured to work in either passive quench or PQAR configurations.

The use of a microcontroller allows for unique identification and access to each APD attached to the circuit through a connected network such as RF wireless and/or Ethernet. 
By using the Zynq-7000 SoC which has space compatibility~\cite{nasa-zynq}, our design is also suitable for nano-satellite applications with size/volume constraints and strict SWAP (size-weight-area-power) budgets. 



The tunable parameters in our design can be varied over a wide range, allowing the same circuit to operate various APDs from different manufacturers without requiring any modifications. 
We demonstrate this versatility by simultaneously operating our system with a commercial APD and a custom fabricated chip-scale APD, both of which have different characteristics in terms of breakdown voltage, optimal over-voltage, etc. 
The flexibility of our system allows it to be applicable to a research setting where multiple device variants are explored and characterized, 
as well as mass-production settings for high-volume wafer-scale testing. 

\section{Design and implementation}
\label{sec:design}
\begin{figure} [!t]
\begin{center}
\begin{tabular}{c} 
\includegraphics[width=0.48\textwidth]{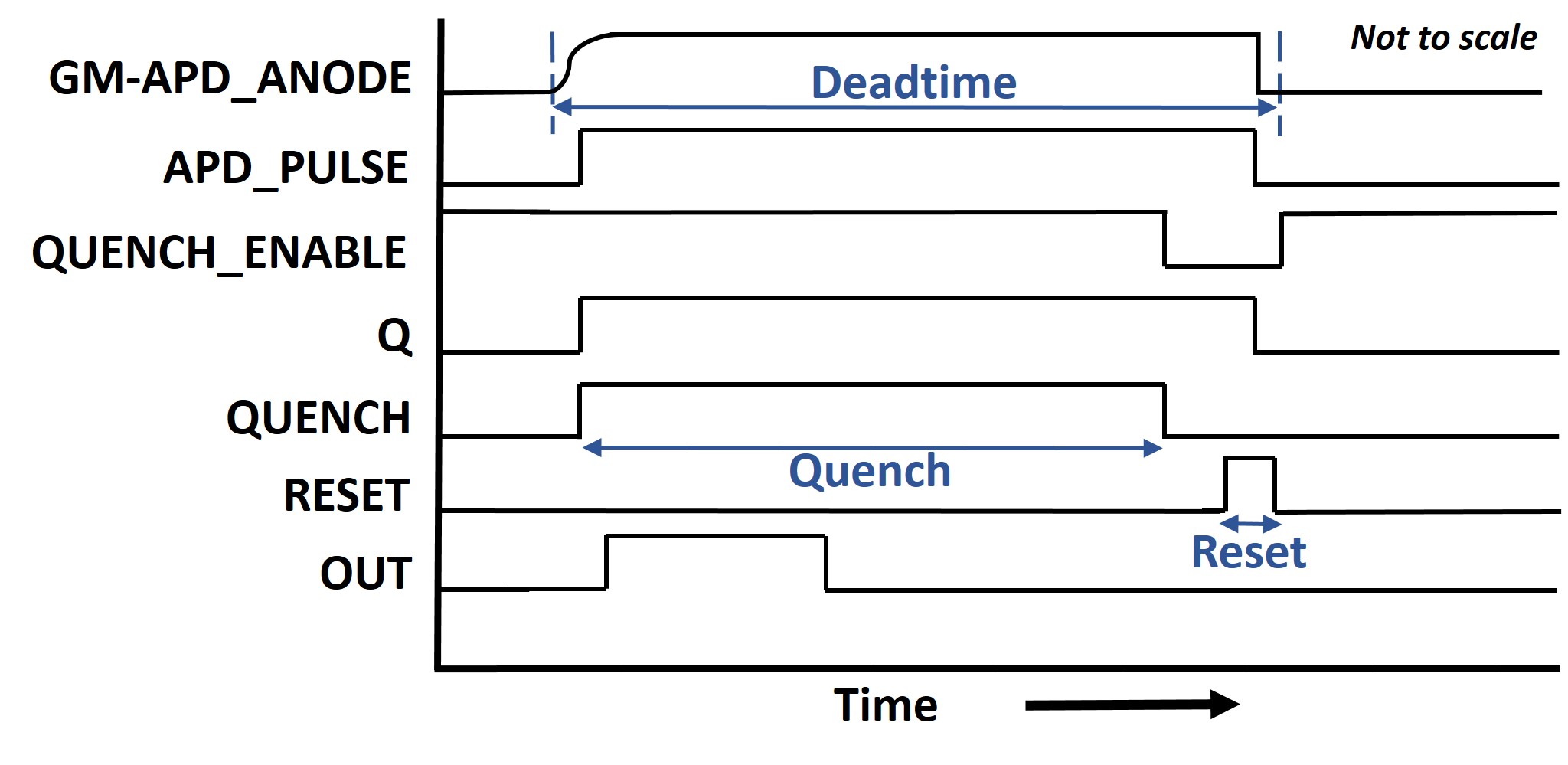}
\end{tabular}
\end{center}
\caption[] 
{ \label{aq-timingdiagram} 
Timing diagram illustrating the active quench operation after the occurrence of an avalanche.}
\end{figure}

The main functional blocks of our system are shown in Figure~\ref{aq-design2}. 
The $ZedBoard$~\cite{zedboard} development kit with a Zynq-7000 SoC is used to implement the corresponding modules on the FPGA and microcontroller. 
A custom PCB is designed using minimal discrete components to connect the SoC with the APD. 
The cathode of the APD is connected to a bias voltage (V\_BIAS) 
while the anode is connected to the ballast resistor $R_B$ in series with the sense resistor $R_S$. 
We use the LTC6752-3 high-speed comparator (COMP) which produces complementary outputs after detecting the avalanche pulse across $R_S$. 
The feedback path for performing the quench comprises of a fast AND gate, P-MOSFET $P_1$ and N-MOSFET $N_1$. 
The APD is then reset through the N-MOSFET $N_2$. 
These MOSFETs are controlled by the signals from the Quench/Reset generator module on the FPGA. The other (inverted) comparator output $\overline{Q}$ (APD\_PULSE) is fed both into the quench/reset logic,
and also into a 24-bit counter to record the number of measured avalanches. 
The values of $R_4$ and $C_1$ are set such that the D-FlipFlop (D-FF) trims the comparator output to 20\,ns-wide pulses (OUT) which are used for characterizing the deadtime and after-pulsing probability. 
On-board high-voltage DC-DC converters (Matsusada) are used to generate a V\_BIAS of between \mbox{0\,-\,500\,V}, while the quench voltage V\_QUENCH is obtained from an external power supply.

\subsection{Operation in active quench configuration}
The active quench operation is illustrated in the timing diagram shown in Figure~\ref{aq-timingdiagram}. 
Upon the absorption of an input photon, the avalanche current in the APD begins to flow. 
The APD is initially passively quenched for a short duration by the ballast resistor ($R_B$) which limits the avalanche current. 
The avalanche current causes the voltage at the APD anode (GM-APD\_ANODE) to rise, 
and consequently the comparator (COMP) input 
exceeds the reference voltage (V\_REF), which is typically set to $\approx$\,17\,mV in our system. 
In response, the comparator toggles the complementary output pulses $Q$ to high and $\overline{Q}$ to low. 
$Q$ is fed back to the AND gate; with the other input of the AND gate (QUENCH\_ENABLE) kept high in the nominal state (before the avalanche occurs), 
the AND gate output (QUENCH) goes high to turn on $N_1$ and subsequently $P_1$. 
As a result, the APD anode voltage reaches the quench voltage (V\_QUENCH), which effectively reduces the voltage across the APD to below its breakdown voltage, and quenches the avalanche. 
This feedback path is kept as short as possible to reduce the response time and after-pulsing probability~\cite{Wayne-AP}.

After the quench duration, the FPGA sets the QUENCH\_ENABLE signal to low, 
which drives the output of the AND gate to low and turns OFF $N_1$ and $P_1$. 
The FPGA then initiates the reset phase by setting the RESET signal high to turn ON $N_2$. 
This bypasses $R_B$ and $R_S$ and brings the GM-APD\_ANODE voltage to 0\,V, and restores the voltage across the APD to above its breakdown voltage. 
The comparator input also drops below V\_REF, which again toggles its output $Q$ and $\overline{Q}$, and consequently 
the FPGA drives the RESET signal to low and turns off $N_2$. 
After the reset duration, the QUENCH\_ENABLE signal is driven high again to ready the system for the next detection event.

The quench and reset durations are configurable, and are set based on the APD characteristics and operational requirements.
To ensure that both $P_1$ and $N_2$ are not ON at the same time, the FPGA also inserts a tiny configurable delay between end of the quench phase and start of the reset phase. 
The total duration from the start of the avalanche to the completion of the reset constitutes the $deadtime$ of the system. 



\begin{figure} [!t]
\begin{center}
\begin{tabular}{c} 
\includegraphics[width=0.9\columnwidth]{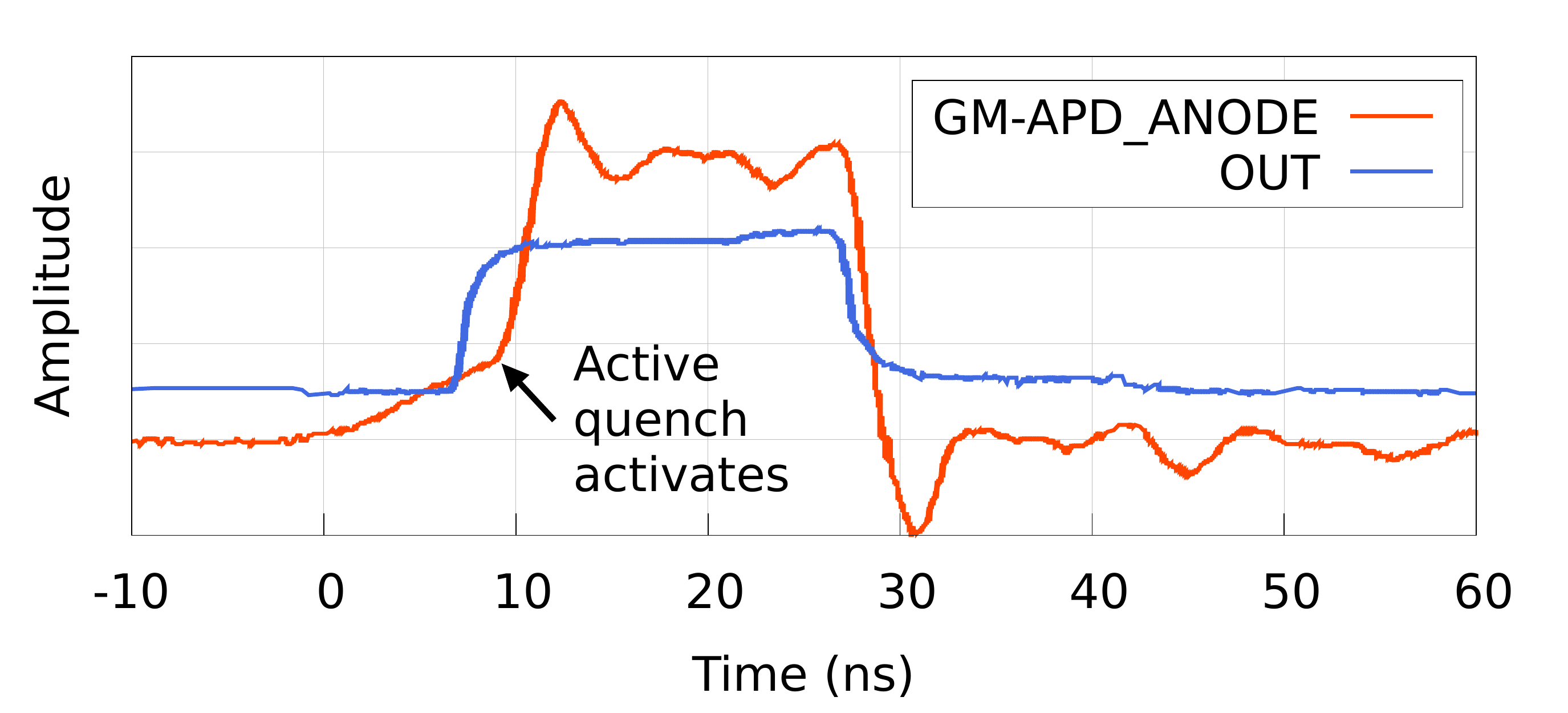}
\end{tabular}
\end{center}
\caption[] 
{ \label{fig:aq-anode} 
Measured oscilloscope waveforms of the GM-APD\_ANODE voltage (5\,V/div)  and the OUT pulse (2\,V/div) 
for an avalanche starting at 0\,ns.
The two traces are offset along the vertical axis for clarity.}
\end{figure}

The operation of our system is verified with a widely-used commercial SAP500-TO8 APD~\cite{sap500} with a typical breakdown voltage of 125\,V at a temperature of 22°C, 
which we bias with an excess voltage of 10V. 
V\_QUENCH is typically set to 15V, 
which allows for APD operation up to an excess voltage of 12\,V. 
Figure~\ref{fig:aq-anode} shows the GM-APD\_ANODE voltage captured with a LeCroy Waverunner 610Zi oscilloscope. 
The active quenching begins $\approx$\,9\,ns after the avalanche starts;
this delay is the sum of propagation delays along the quench feedback path: through the comparator, AND gate and the switch ON delays for $N_1$ and $P_1$ (see Figure \ref{aq-design2}).
After the quench is activated, the GM-APD\_ANODE voltage rises rapidly to V\_QUENCH within 2\,ns. 
The OUT signal is delayed by $\approx$\,7\,ns from the start of the avalanche due to the propagation delays through the comparator, inverting buffer and D-Flipflop. 

The minimum quench duration is determined by the propagation delay of the APD\_PULSE signal from the comparator through the FPGA logic to drive QUENCH\_ENABLE signal and AND gate output to low; 
in our case this is 10\,ns, with a physical separation of $\approx$\,15\,cm between the APD head and the SoC. 
The delay between the quench and reset phases is typically set to 5\,ns to account for switch OFF delays of $N_1$ and $P_1$. 
Accordingly, in Figure \ref{fig:aq-anode}, we observe that the reset is activated 15\,ns after GM-APD\_ANODE reaches V\_QUENCH. 
The reset pulse duration is kept as small as possible and is typically between 5\,--\,10\,ns depending on the value of V\_QUENCH. 
Table \ref{tab:reconfig-param} shows the list of configurable parameters in our design along with the range of supported values. 
The FPGA module for quench/reset generation runs on an internal 200\,MHz clock, which allows for the control of the timing parameters with a resolution of 5\,ns.
\begin{table}[!t]
\caption{Reconfigurable parameters and their range of supported values} 
\label{tab:reconfig-param}
\begin{center}       
\begin{tabular}{|l|c|c|c|}
\hline
\rule[-1ex]{0pt}{3.5ex}  & Min. & Max. & Units \\
\hline
\rule[-1ex]{0pt}{3.5ex} Quench duration & 10 & 1000 & ns  \\
\hline
\rule[-1ex]{0pt}{3.5ex} Reset duration & 5 & 1000 & ns  \\
\hline
\rule[-1ex]{0pt}{3.5ex} Deadtime & 35 & 1000 & ns  \\
\hline
\rule[-1ex]{0pt}{3.5ex} V\_QUENCH & 0 & 30 & V  \\
\hline
\rule[-1ex]{0pt}{3.5ex} V\_BIAS & 0 & 500 & V  \\
\hline 
\end{tabular}
\end{center}
\end{table}
\subsection{Operation in PQAR and passive quench configurations}
Our system can also be configured to operate in either a passive quench active reset (PQAR) mode or passive quench mode.
In the PQAR mode, the FPGA is configured to keep the QUENCH\_ENABLE signal low at all times. 
This disables the AND gate and the QUENCH signal is never triggered, keeping $N_1$ and $P_1$ always OFF (see Figure \ref{aq-design2}). 
Thus when an avalanche occurs, the APD is quenched passively through ballast resistor $R_B$ for a configurable quench
duration, 
after which the FPGA drives only the RESET signal to high and turns ON $N_2$. 
This restores the reverse bias voltage across the APD to V\_BIAS. 
After a configurable reset duration the RESET signal is set to low again to turn OFF $N_2$. 

In the passive quench mode, the FPGA is configured to keep both the QUENCH\_ENABLE and RESET signals low at all times. 
As a result $P_1,\ N_1$ and $N_2$ are never turned ON and so the APD is both quenched and reset passively. 

In both modes the comparator still continues to detect APD avalanches, and toggles the OUT and APD\_PULSE signals which can be counted by the FPGA.
\section{Performance characterization with commercial APD}
We characterize our system with the SAP500-T8 APD which has a built-in TEC (Thermo-electric cooler) to vary the temperature. 
We present the lowest achievable values for deadtime and after-pulsing probability at an excess voltage of 10\,V for a range of temperatures.


\subsection{Deadtime}
\begin{figure} [!t]
\begin{center}
\includegraphics[width=0.9\columnwidth]{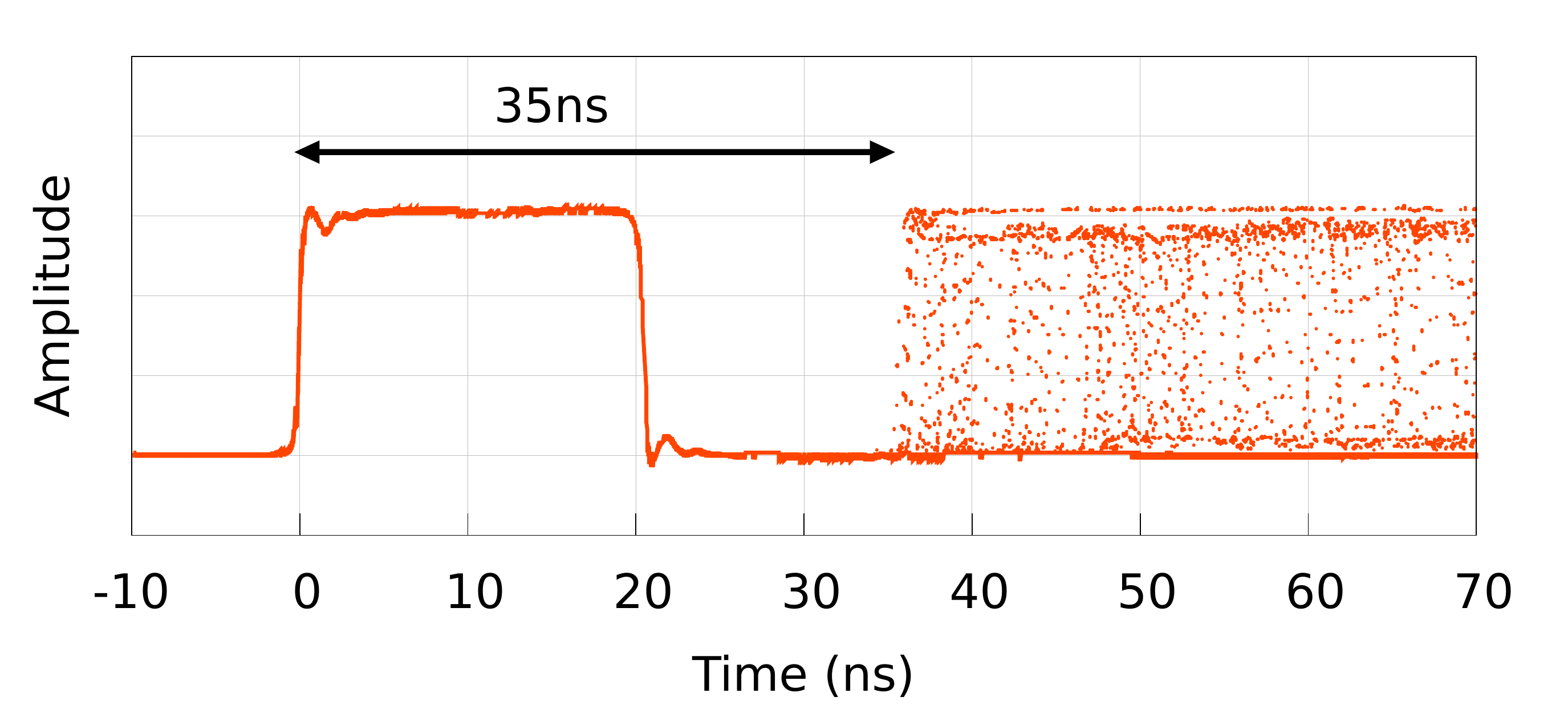}
\end{center}
\caption[] 
{ \label{fig:deadtime} 
A deadtime of 35\,ns is observed with accumulated OUT pulses (1\,V/div) captured on the oscilloscope in persistence mode. 
The APD is measuring background light at a count rate of $\approx$\,66\,kcps.}
\end{figure}

Once an avalanche is detected, the APD cannot detect another avalanche until the first avalanche is completely quenched and the APD is reset to its original state. This constitutes the deadtime as shown in Figure~\ref{aq-timingdiagram}. 
We capture the accumulated OUT pulses on the oscilloscope in persistence mode. It can be observed in Figure~\ref{fig:deadtime} that the deadtime (i.e. the time interval between the rising edge of the first pulse and the rising edge of the earliest subsequent pulse) is 35\,ns 
which, in principle, allows for detection rates of $>$28\,Mcps with our system. 
\subsection{After-pulsing probability and temperature}
\begin{figure}[!t]
\centering
\includegraphics[width=0.99\columnwidth]{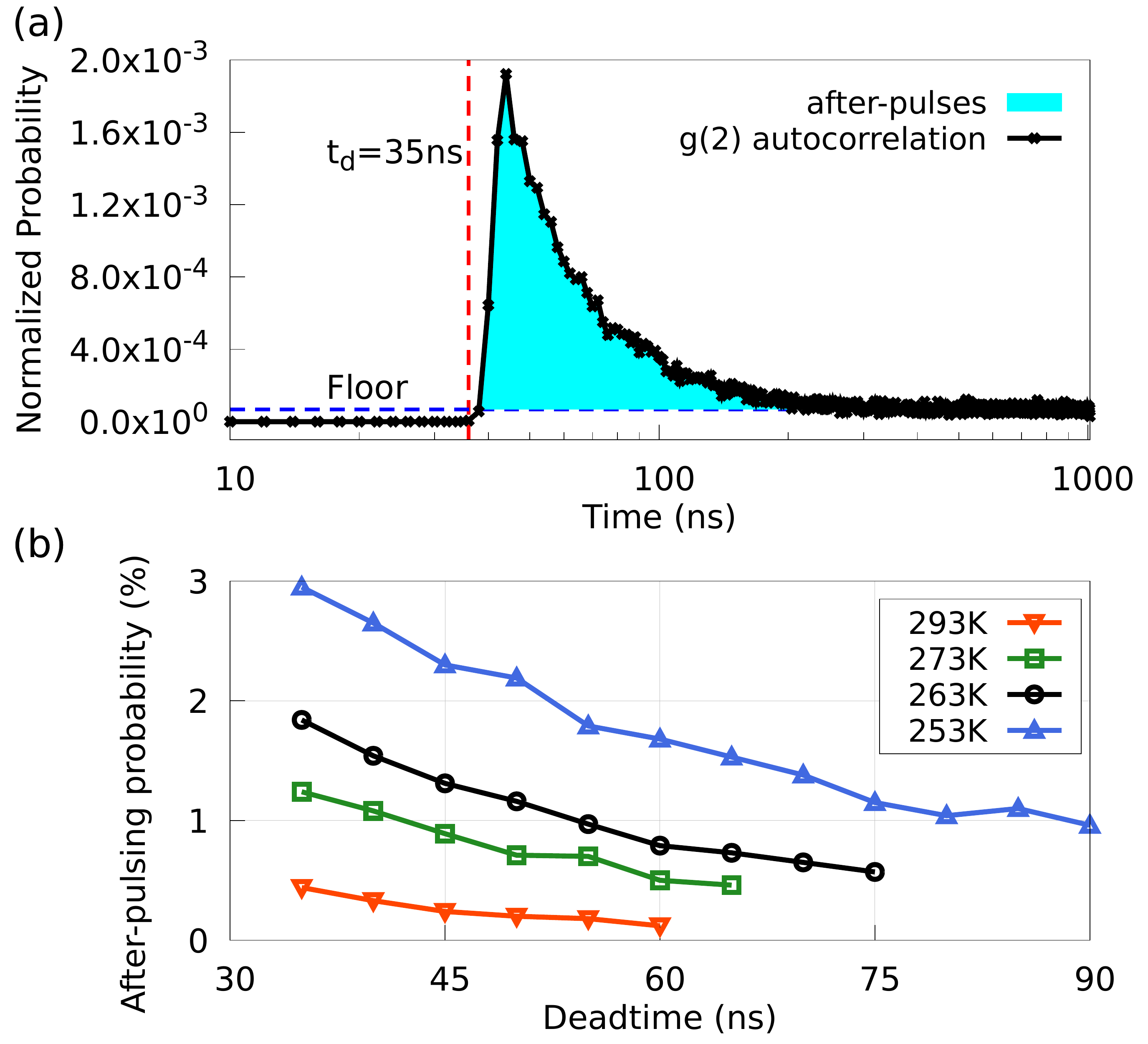}
\caption{After-pulsing in the SAP500-T8 APD, operating at an excess voltage of 10\,V and a background count rate of 30\,kcps. 
(a) The second order auto-correlation $g^{(2)}$ of OUT pulses with a time bin width of 2\,ns. 
The number of after-pulses (highlighted) is higher immediately after the deadtime $t_d$ of 35\,ns (vertical red line) from the onset of the avalanche, and tapers off to reach a floor (horizontal blue line) after $\approx$\,200\,ns. 
The APD is at a temperature of 253\,K.
(b) Variation of after-pulsing probability at different temperatures and deadtime. }
\label{fig:afterpulse}
\end{figure}

It is beneficial to operate the APD at lower temperatures as it reduces the dark counts (occurring due to thermal excitation), 
however this also increases the after-pulsing probability. 
The after-pulsing probability also increases when the deadtime of the circuit is reduced. 
We measure the after-pulsing probability for the SAP500-T8 using our system for different deadtimes and temperatures. 
This is done by computing the second order auto-correlation function $g^{(2)}$ of the OUT pulses using an external time-tagger module (with 2\,ns time resolution). 
Figure~\ref{fig:afterpulse}(a) shows our results for a deadtime ($t_d$) of 35\,ns, APD temperature of 253\,K (-20°C), over-voltage of 10\,V, and a background rate of $\approx$\,30\,kcps. 
The APD output is recorded for 10\,s which gives more than 300,000 data samples ($C_{Total}$) of output pulse arrival times. 
The $g^{(2)}$ auto-correlation is then computed with a time resolution of 2\,ns. 
In Figure~\ref{fig:afterpulse}(a) it can be seen that there is a high degree of correlation of arrival times immediately after the deadtime; 
after reaching a peak, the correlations drop exponentially until they reach a very low background `floor' level.
The correlations above the background level constitute the after-pulses as highlighted in the graph. 

Quantitatively, the after-pulse probability $P_{ap}$~\cite{Ceccarelli2019} can be defined as:
\begin{equation}
    P_{ap} = \frac{\sum_{\tau=t_d}^{+\infty}(C_{\tau}-floor)}{C_{Total}}
    \label{eq:afterpulse} 
\end{equation}
where $C_{Total}$ is the total number of events and $C_{\tau}$ is the count value of each time bin at time $\tau$ after the avalanche pulse. 
The after-pulsing probability of the SAP500-T8 APD obtained with our system is shown in Figure \ref{fig:afterpulse}(b) for four operating temperatures (-20°C, -10°C, 0°C and +20°C). 
The deadtime is varied in steps of 5\,ns by changing only the quench duration while keeping all other parameters constant. 
In our analysis,
we consider $t_d$ as the time bin of the first observed correlation event, 
and 10$\,\upmu$s as the upper limit for $\tau$. It is observed that $P_{ap}$ at the lowest temperature (-20°C) and lowest deadtime (35\,ns) is 2.95\,$\pm$\,0.08\%, and remains below this value for all other settings of deadtime and temperature. 
For most practical applications such as QKD, it is important to keep the after-pulsing probability below 5\% \cite{Stipcevic17, Ceccarelli2019}.

\section{Integration with custom chip-scale APD}

After validating the performance of our active quench system with the commercial APD, 
we proceeded to work with custom-fabricated chip-scale APDs~\cite{imreAPD21}. 
These Si waveguide APDs are integrated with on-chip photonic waveguides, and are interfaced with the active quench system using RF probes.
We included a second identical channel in our system so that both the SAP500-T8 and custom APD can be simultaneously connected and operated in parallel. 
Both APDs are controlled by only one SoC through replication of the quench/reset generator and the counter modules on the FPGA. 
The SAP500 was operated at room temperature while the temperature of the chip-scale APD was varied between -20°C and +50°C using a peltier cooler.
To rule out the presence of any stray light in the setup, the count rate of the SAP500 was monitored closely during the experiments to ensure that the dark counts always remained at $\approx$\,20\,kcps, which is the expected value for this APD at room temperature when operated at an excess voltage of 10\,V.
As the breakdown voltages of the chip-scale APDs vary from 16\,--\,25\,V, V\_BIAS is directly obtained from a bench-top power supply. 
To prevent the breakdown voltage of the chip-scale APDs from drifting~\cite{imreAPD21}, V\_QUENCH was maintained at 2V above V\_BIAS. 
Considering the higher V\_QUENCH values (18\,--\,27\,V), the durations of quench, reset and deadtime for both channels were set conservatively higher so as to provide sufficient time for the MOSFETs to switch ON/OFF.
Thus, for both APDs, the quench duration was set to 25\,ns and the reset duration to 15\,ns, with a delay of 10\,ns between quench and reset. 
This gives a total deadtime of $\approx$\,65\,ns (inclusive of the initial response time and all other delays) for all measurements. 
V\_QUENCH and V\_BIAS of the chip-scale APD were changed at run-time using scripts running on a laptop PC, while the other parameters listed in Table~\ref{tab:reconfig-param} remained fixed.

\begin{figure} [!t]
\begin{center}
\begin{tabular}{c} 
\includegraphics[width=0.95\columnwidth]{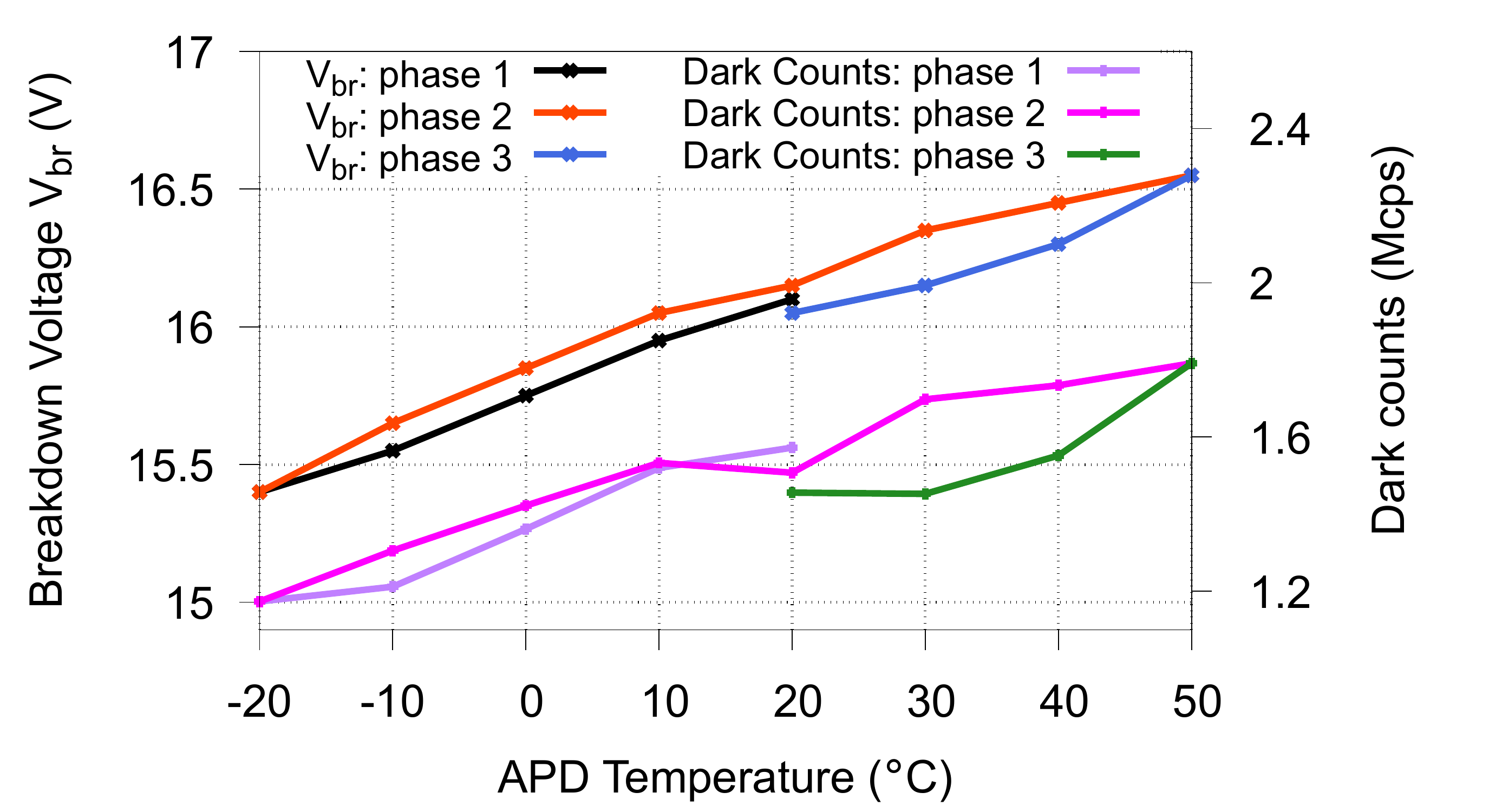}
\end{tabular}
\end{center}
\caption[] 
{ \label{fig:astar-data} 
Variation of breakdown voltage and dark counts with temperature for a chip-scale APD.
The data was obtained in three consecutive temperature sweeps, during which the APD was first cooled down to -20°C, then warmed up to +50°C, then cooled to +20°C. 
The quench voltage was set at 2\,V above the bias voltage.}
\end{figure} 
For commercial APDs, it is well understood that the breakdown voltage increases with temperature (and vice-versa), 
and does not exhibit hysteresis during temperature cycling. 
To investigate if the same behaviour can be observed for a chip-scale APD, we  measured its breakdown voltage while changing the temperature in three phases:  
it was first cooled from +20°C to -20°C, then warmed up from -20°C to +50°C and again cooled down from +50°C to +20°C. 
The temperature was changed in steps of 10°C. 

At each temperature step, V\_BIAS was initially set to a value well below the breakdown voltage and the number of detected counts was verified to be 0. 
V\_BIAS was then increased in steps of 0.05\,V every 2 seconds until we started measuring counts, which indicated that V\_BIAS has reached the breakdown voltage. 
V\_BIAS was then kept constant and the dark counts were recorded. 
Before each new measurement the APD was reset by switching off V\_BIAS. 

Figure \ref{fig:astar-data} shows the variation in dark count rate and breakdown voltage with temperature for the chip-scale APD. It can be observed that the results for the chip-scale APD follows the same trend as the commercial device.

\subsection{Advantage of using SoC based active quenching for testing chip-scale APDs}
For our preliminary tests, only the dark counts and breakdown voltage of one chip-scale APD have been presented. 
However, the system can be easily programmed to scale up the tests to measure multiple APD devices at different operating conditions,
e.g. the characterization of after-pulsing probability of each APD for at least 10 different deadtimes ($t_d$), 5 different temperatures ($T$) and 3 different excess voltages ($V_e$), which amounts to a minimum of 150 test cases per APD ($t_d \times T \times V_e$). 
When fabricated on a wafer-scale, there can be hundreds of device variants differing in device length, width, doping concentrations, etc. Characterizing all of them would then require the active quench system to be re-configured tens of thousands of times to execute all the test cases. 
Using fixed active quench circuits with hard-wired components for such testing is clearly impractical since the required manual intervention for re-configuration would be very laborious and error prone. 
Instead, our SoC-based active quench system paves the way for automated script-based re-configuration and testing that can be scaled up, leading to significant cost and time savings.

\section{Conclusion and future work}
We have described a hybrid design for a re-configurable active quench system which can be used to characterize multiple types of APDs. 
We have implemented the design with readily available components and characterized it with a SAP500-T8 commercial APD cooled to -20°C, and achieved a deadtime of 35\,ns while limiting the after-pulsing probability to $\approx$\,3\%. 
We have also integrated our active quench system directly with custom chip-scale APDs, and presented preliminary characterization results. 
Our active quench system can applied to many applications such as quantum information processing, random number generators, time-correlated single photon counting (TCSPC), and range measurement applications such as LIDAR. 
Looking ahead, the deadtime of the quenching system can be reduced further by using faster MOSFETs which will increase the detection rate and efficiency.
The SoC firmware can also be upgraded to directly perform data post-processing, e.g. computing after-pulsing probability, without additional equipment. 

\section{Acknowledgements}
This research is supported by the National Research Foundation, Singapore under its Central Gap Fund (NRF2018-NRFCG001-001)
and QEP2.0 programme (NRF2021-QEP2-03-P04),
as well as the Agency for Science, Technology and Research, Singapore (\#21709).

\bibliography{Bib-ReconfAQ} 
\bibliographystyle{IEEEtran}

\end{document}